\documentstyle[eqsecnum,pre,preprint,aps,epsfig,amsmath,amssymb]{revtex}

\tightenlines

\newcommand{\be}{\begin{equation}}
\newcommand{\ee}{\end{equation}}
\newcommand{\ba}{\begin{eqnarray}}
\newcommand{\ea}{\end{eqnarray}}
\newcommand{\bq}{\begin{equation}}
\newcommand{\eq}{\end{equation}}
\newcommand{\bqa}{\begin{eqnarray}}
\newcommand{\eqa}{\end{eqnarray}}
\newcommand{\ben}{\begin{enumerate}}
\newcommand{\een}{\end{enumerate}}
\newcommand{\bc}{\begin{center}}
\newcommand{\ec}{\end{center}}
\newcommand{\bqb}{\begin{eqnarray*}}
\newcommand{\eqb}{\end{eqnarray*}}

\newcommand{\mttbar}{M_{t\overline t}}

\begin{document}

\draft
\preprint{PM/05-20}

\title{\vspace{1cm}
Weak Interaction Sum Rules for Polarized $t\overline{t}$ 
Production at LHC
\footnote{Partially supported by EU contract HPRN-CT-2000-00149}}
\author{M. Beccaria$^{a,b}$,
F.M. Renard$^c$ and C. Verzegnassi$^{d, e}$ \\
\vspace{0.4cm}
}

\address{
S$^a$Dipartimento di Fisica, Universit\`a di
Lecce \\
Via Arnesano, 73100 Lecce, Italy.\\
\vspace{0.2cm}
$^b$INFN, Sezione di Lecce\\
\vspace{0.2cm}
$^c$ Laboratoire de Physique Th\'{e}orique et Astroparticules,
UMR 5207\\
Universit\'{e} Montpellier II,
 F-34095 Montpellier Cedex 5.\hspace{2.2cm}\\
\vspace{0.2cm}
$^d$
Dipartimento di Fisica Teorica, Universit\`a di Trieste, \\
Strada Costiera
 14, Miramare (Trieste) \\
\vspace{0.2cm}
$^e$ INFN, Sezione di Trieste\\
}

\maketitle

\begin{abstract}
We consider two polarization asymmetries in the process of top-antitop 
production at LHC. We show that the theoretical predictions for these two 
quantities, at the strong and electroweak partonic one-loop level, are free of 
QCD and QED effects. At this perturbative level we derive two sum rules, that 
relate measurable quantities of top-antitop production to genuinely weak 
inputs. This would allow to perform two independent tests of the candidate 
theoretical model, with a precision that will be fixed by the future 
experimental accuracies of the different polarization measurements. A 
tentative quantitative illustration of this statement for a specific MSSM scenario
is enclosed, and a generalization to include two other future 
realistic measurements is also proposed.
\end{abstract}

\pacs{PACS numbers: 12.15.-y, 12.15.Lk, 13.75.Cs, 14.80.Ly}

\section{Introduction}

It is by now generally accepted that top-antitop production at LHC will be a 
potential source of very relevant theoretical information, both within the 
Standard Model description and within other possible new physics schemes~\cite{CERNYB}. 
In this spirit, the possibility 
of performing high precision measurements of the process has been considered 
with some attention in the literature. The general difficulty that one 
encounters in this case is that two different sources of uncertainty have to 
be carefully taken into account. Actually, on top of the experimental 
statistical and systematic errors, a not negligible theoretical uncertainty 
of, essentially, QCD origin usually affects the available theoretical 
predictions. In the case of unpolarized cross sections, the size of the 
theoretical QCD uncertainty is at the moment estimated to be of, approximately, 
a relative twelve percent~\cite{CERNYB}. This value is not much smaller 
than that of the corresponding overall experimental error, for which a recent 
preliminary estimate has derived a rather conservative upper limit of 
approximately twenty percent~\cite{Beccaria:2004sx}. Assuming that this 
experimental limit will be reduced to a ~ten percent level by more dedicated 
future efforts (which appears a reasonable aim in the discussions of~\cite{CERNYB}), 
one expects a future situation where the experimental and the QCD 
uncertainties would be of, roughly, equal $\sim$ ten percent size for the 
production of unpolarized top-antitop pairs.

If the theoretical model to be tested is one of supersymmetric kind, like for 
instance the MSSM, and the purpose of the investigation is that of measuring 
weak supersymmetric virtual effects at the one-loop level,assuming a 
preliminary direct Supersymmetry discovery, the previous discussion shows that 
it might be difficult to identify virtual effects if they were not beyond the 
ten percent limit. For special scenarios of light Supersymmetry and large $\tan\beta$
values one might well find  the case of such large ($\sim$ twenty percent) 
effects, as exhaustively discussed in~\cite{Beccaria:2004sx},
where a particular approach was proposed, based on measurements of the slope of the final invariant mass 
distribution $d\sigma/d\mttbar$. But in a less "friendly" supersymmetric scenario, 
with possibly smaller virtual SUSY effects, a different search would be 
requested, and the size of the QCD uncertainty might add an extra difficulty 
to the analysis. Clearly, this difficulty would be "substantially" reduced if 
a different experimental quantity could be measured that turned out to be, 
within certain reasonable assumptions, "substantially" less sensitive to QCD 
effects.

The aim of this paper is precisely that of proposing the measurements of two
observables of the process that would meet the previous request. Both 
quantities are certain polarization asymmetries, and therefore the needed 
measurements would be those of the final top-antitop helicities. Although the 
topics is not a new one, and several excellent papers exist in the literature 
devoted to a description of the theoretical properties of the top-antitop 
polarized cross sections, we shall devote the next Section 2 to a brief 
summary of those features that are essential for our approach. In Section 3 a 
tentative quantitative illustration of the possible outcomes of our proposal 
will be also briefly proposed.

\section{Polarized \lowercase{$t\overline{t}$} production}

We start by considering top-antitop production at the Born level, with a final
invariant mass $\mttbar$ (that coincides, at Born level, with the initial 
partons c.m. energy $\sqrt{s}$). A very special feature of the process, not
shared by any other light (u,d,s,c,b) quark pair production at the 
corresponding energies, is that both top and antitop can be separately 
produced in two different helicity states. This is a consequence of the large 
top mass, that generates two "unconventional" helicity pairs (i.e. not of the 
massless quark kind) accompanied by typical $m_t^2/s$ factors. In our 
notations, that are essentially similar to those of~\cite{Beccaria:2004sx}, we shall 
label states by their chirality at high energy. Thus, the two 
combinations (L,L) and (R,R) would be in our notation the pairs with opposite 
helicities, while (L,R) and (R,L) would be those with equal helicities.

At LHC, the dominant production mechanism is due to an initial gluon-gluon 
pair. As already discussed in previous articles (see e.g. \cite{Stelzer:1995gc}), 
the two equal (LR and RL) and opposite (LL and RR) helicity 
pairs are in this case orthogonal, in the sense that equal helicities 
production dominates at low energies, opposite helicities production dominates 
at high energies. For an initial quark-antiquark pair, that represents the 
leading mechanism at Tevatron, the dominant production would be in each case 
that of opposite helicity pairs,less strongly at at low energies and more 
strongly at high energies, where it would reproduce the LHC situation. To get 
a more quantitative description, we have computed, at Born level, the overall 
LHC (LR +RL) and (LL + RR) invariant  mass distributions  
$d\sigma_{LL}/d\mttbar = d\sigma_{RR}/d\mttbar$ and 
$d\sigma_{LR}/d\mttbar = d\sigma_{RL}/d\mttbar$. 
With this aim, we have started from the Born expressions of the differential 
cross sections at partonic level that are, for the initial gluon-gluon state:
\bqa
\frac{d\sigma^{Born}(gg\to t_L\bar t_L)}{d\cos\vartheta} &=&
\frac{d\sigma^{Born}(gg\to t_R\bar t_R)}{d\cos\vartheta}=\nonumber\\
&&
{\pi\beta^3\alpha^2_s\sin^2\vartheta(1+\cos^2\vartheta)
(7+9\beta^2\cos^2\vartheta)\over192 s(1-\beta^2\cos^2\vartheta)^2}.
\eqa
\bqa
{d\sigma^{Born}(gg\to t_L\bar t_R)\over d\cos\vartheta}&=&
{d\sigma^{Born}(gg\to t_R\bar t_L)\over d\cos\vartheta}=\nonumber\\
&&
{\pi\beta\alpha^2_sm^2_t(1+\beta^2(1+\sin^4\vartheta))
(7+9\beta^2\cos^2\vartheta)\over48 s^2(1-\beta^2\cos^2\vartheta)^2}
\eqa
where $\beta=\sqrt{1-4m^2_t/s}$.

For initial quark-antiquark state we have:
\bq
\frac{d\sigma^{Born}(q\overline q\to t_L\bar t_L)}{d\cos\vartheta} =
\frac{d\sigma^{Born}(q\overline q\to t_R\bar t_R)}{d\cos\vartheta}= 
{\pi\alpha^2_s\beta\over18 s}
(1+\cos^2\vartheta)
\eq

\bq
{d\sigma^{Born}(q\overline q\to t_L\bar t_R)\over d\cos\vartheta} = 
{d\sigma^{Born}(q\overline q\to t_R\bar t_L)\over d\cos\vartheta} =
{2\pi\alpha^2_s\beta m^2_t\over9 s^2}\sin^2\vartheta 
\eq

Starting from these expressions, and working systematically at Born level, we 
have next computed the distributions ($a$ and $b$ can be $L$ or $R$)
\bqa
\label{eq:hadroniclevel}
{d\sigma(PP\to t_a\bar t_b+...)\over ds}&=&
{1\over S}~\int^{\cos\vartheta_{max}}_{\cos\vartheta_{min}}
d\cos\vartheta~[~\sum_{ij}~L_{ij}(\tau, \cos\vartheta)
{d\sigma_{ij\to  t_a\bar t_b}\over d\cos\vartheta}(s)~]
\eqa
\noindent
where $\tau={s\over S}$, and $(ij)$ represent all initial $q\bar q$ pairs with 
$q=u,d,s,c,b$ and the initial $gg$ pairs, with the corresponding
luminosities

\bq
L_{ij}(\tau, \cos\vartheta)={1\over1+\delta_{ij}}
\int^{\bar y_{max}}_{\bar y_{min}}d\bar y~ 
~[~ i(x) j({\tau\over x})+j(x)i({\tau\over x})~]
\eq
\noindent
where $S$ is the total proton-proton c.m. energy, and 
$i(x)$ the distributions of the parton $i$ inside the proton
with a momentum fraction, $x={\sqrt{s\over S}}~e^{\bar y}$, related to the rapidity
$\bar y$ of the $t\bar t$ system~\cite{QCDcoll}.
The parton distribution functions are the latest MRST set~\cite{wlumi}.
The limits of integrations for $\bar y$ can be written
\bqa
&&\bar y_{max}=\max\{0, \min\{Y-{1\over2}\log\chi,~Y+{1\over2}\log\chi,
~-\log(\sqrt{\tau})\}\}\nonumber\\
&&
\bar y_{min}= - \bar y_{max}
\eqa
\noindent
where the maximal rapidity is $Y=2$. The  
quantity $\chi$ is related to the scattering angle
in the $t\bar t$ c.m. by the relation $\chi=(1+\beta\cos\vartheta)/(1-\beta\cos\vartheta)$ 
where $\beta=\sqrt{1-4m^2_t/s}$. The integration limits are  
$\cos\vartheta_{min,max}=\mp\sqrt{1-4p^2_{T,min}/ s}$ expressed in terms of
the chosen value for $p_{T,min} = 50$ GeV.\\

\begin{figure}
\centering
\epsfig{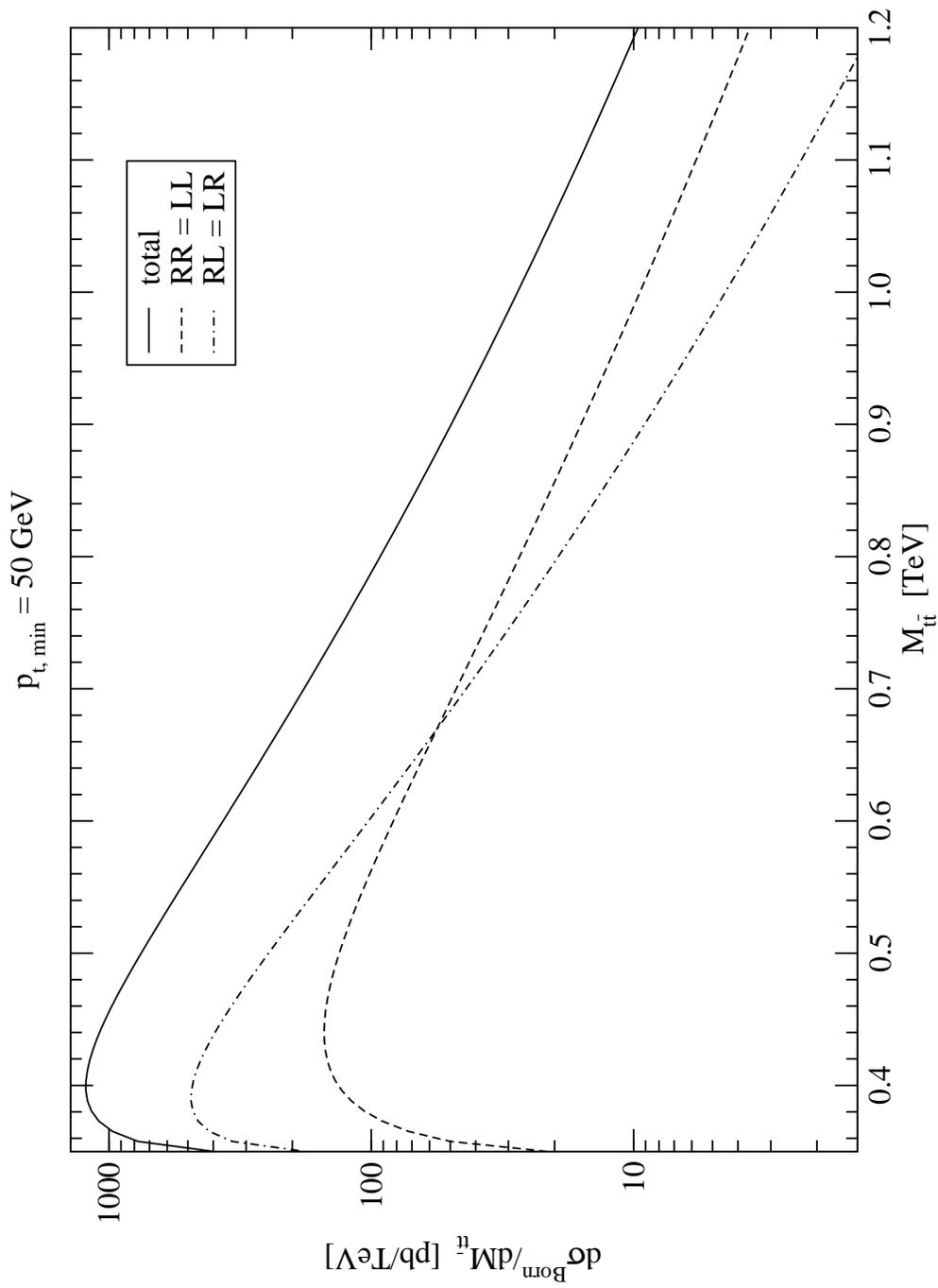}
\vspace{1.5cm}
\caption{
Born value of the distributions $d\sigma_{ab}/d\mttbar$ at hadronic level
for the four possible helicity combinations LL, LR, RL, RR.
}
\label{fig:born}
\end{figure}

The resulting curves are depicted in Fig.~\ref{fig:born}. The 
upper limit for $\mttbar$ has been taken at $\sim 1.2$ TeV, where the recent analysis 
of~\cite{Beccaria:2004sx} shows that one can still expect a reasonable number of events. One 
sees from our analysis, that reproduces correctly the Stelzer-Willenbrock 
curve~\cite{Stelzer:1995gc} for $\mttbar \lesssim 0.8$ TeV (the limit of that calculation), that the 
asymptotically leading opposite helicities production starts being really 
more, but not "much" more, relevant exactly at that energy, becoming about three times
larger than the "competitor" production at $\mttbar \simeq 1.2$ TeV, and 
being definitely depressed in the region around threshold and below $\sim 500$ GeV.
Thus, in the overall realistic LHC energy range for top-antitop production, 
the "asymptotically depressed" equal helicity production must be carefully 
taken into account for an accurate theoretical description of the process.

Until now, our analysis has been limited to the consideration of the 
Born level description. For a realistic analysis, one must now 
move to the next one-loop level. This has to be done both for the QCD and for 
the electroweak virtual effects. Before making this effort, though, we want to 
make  a remark that will turn out to be essential for our approach. This is 
related to the diagrams that contribute, at Born level, the different helicity 
productions. As already pointed out in~\cite{Beccaria:2004sx}, the production of (LL,RR) from a two gluon state is 
only due to the same $t$ and $u$ channel exchanges, while (LR and RL) are coming 
from $t$, $u$ and $s$ channel diagrams (thus canceling for $\mttbar \gg m_t$). Adding the 
small quark-antiquark contribution, that only comes from an $s$ channel gluon 
exchange , one concludes that the Born diagrams for (LL) and (RR) production 
are the same, and are different from those that determine the (LR) and (RL) 
case. Moving now to the one-loop partonic level, we shall write therefore in 
full generality:

\be
\frac{d\sigma^{\rm 1\ loop}_{ab}}{d\cos\vartheta} = 
\frac{d\sigma^{\rm Born}_{ab}}{d\cos\vartheta} (1 + \alpha_s F_{ab}^{\rm QCD} + \alpha F_{ab}^{\rm EW})
\ee

A clear statement must be made at this point. Our treatment of the 
perturbative expansion is rigorously performed at the one-loop level, i.e. 
neglecting extra terms of order $\alpha_s^2$, $\alpha^2$, and $\alpha_s\alpha$. 
In particular, considering the latter ones would also imply  mixed 
strong-electroweak effects at the two-loop level, well beyond the theoretical 
purposes at LHC. Having made this statement, we can now observe that the 
one-loop QCD corrections $F_{LL}^{\rm QCD}$ and $F_{RR}^{\rm QCD}$ are necessarily equal, since 
they come from gluon additions to the SAME set of Born diagrams, and the 
gluons do not distinguish the top or the antitop  helicities. In full analogy, 
we conclude that $F_{LR}^{\rm QCD}$ must be equal to $F_{RL}^{\rm QCD}$. 
These conclusions do not 
apply to the electroweak functions $F_{ab}^{\rm EW}$, since weak 
exchanges discriminate in general L from R. In full generality, we are 
therefore led to the statement that
\ba
\label{eq:diff1}
d\sigma^{\rm 1 \ loop}_{LL}-d\sigma^{\rm 1\ loop}_{RR} &=& d\sigma^{\rm Born}_{LL=RR}\ \alpha\ (F_{LL}^{\rm EW} - F_{RR}^{\rm EW}) \\ 
\label{eq:diff2}
d\sigma^{\rm 1 \ loop}_{LR}-d\sigma^{\rm 1\ loop}_{RL} &=& d\sigma^{\rm Born}_{LR=RL}\ \alpha\  (F_{LR}^{\rm EW} - F_{RL}^{\rm EW}) 
\ea
i.e. the differences of the previous cross sections are, at the one-loop 
level, free of QCD effects. In fact, one can be even more stringent since, for the 
same reasons that eliminate the virtual gluon corrections, also the virtual 
photon corrections are washed out in the differences. In other words, the two 
quantities Eqs.~(\ref{eq:diff1}-\ref{eq:diff2}) are, at the one loop level, free of both QCD and QED
virtual effects. 

Before continuing our analysis, we feel that it is opportune at this point to 
make two extra remarks. The first one is that, for what concerns the "QCD-QED 
freedom" of the two Eqs.~(\ref{eq:diff1}-\ref{eq:diff2}), our conclusion is only valid if one 
considers, as we did, the differences of the cross sections. For the sums of 
the polarized quantities, the cancellation of the QCD (and QED) virtual 
effects would not be obtained, so that in those cases the calculation of those 
terms should be rigorously performed. One might imagine to remove this 
difficulty by considering
"conventional" polarization asymmetries, defined as the ratios of the 
differences (\ref{eq:diff1}-\ref{eq:diff2}) to the corresponding sums. This possibility, 
first proposed by Kao and Wackeroth~\cite{Kao:1999kj} and more recently reconsidered in~\cite{Beccaria:2004sx} 
for the (LL,RR) case, leads to the definition of invariant mass 
distributions of the kind :

\be
A_{LL,RR}(\mttbar) = \frac{d\sigma_{LL}-d\sigma_{RR}}{d\sigma_{LL}+d\sigma_{RR}}
\ee  
and analogously: 
\be
A_{LR,RL}(\mttbar) = \frac{d\sigma_{LR}-d\sigma_{RL}}{d\sigma_{LR}+d\sigma_{RL}}
\ee
\noindent
where the various components should be computed at the $pp$ level as in Eq.~(\ref{eq:hadroniclevel}).
At the one-loop level, Eqs.(2.11) and (2.12) would also actually be QCD and QED free. 
The reason why we believe that these asymmetries might not be the best choice 
for an experimental test is related to the second remark that we anticipated. 
This is the statement that the theoretical results of our investigation must 
be compared with realistic experimental measurements.In this sense, a problem 
that we can imagine for the measurement of the invariant mass distribution of 
general polarized cross sections is that a precise determination of the top  
(antitop) helicity usually generates a loss of precision for the value of the 
invariant mass itself\footnote{We thank Stan Bentvelsen for an illuminating 
discussion on this point}~\cite{CERNYB}. This problem could be avoided if one considered, 
rather than the invariant mass distributions, the integration of the polarized 
cross sections, the integral being performed between a lower value and an 
upper value of the invariant mass to be conveniently chosen. In this spirit, 
we shall therefore consider from now on as potential measurable candidates the 
following quantities (${\cal L}_{int}$ being the integrated luminosity):

\be
\label{eq:number}
N_{ab}(s_1, s_2)= {\cal L}_{int} \int_{s_1}^{s_2}\frac{d\sigma_{ab}}{ds}\ ds
\ee

Note that we used for simplicity the initial parton c.m. energy 
$\sqrt{s}$, since the difference at one-loop between $\sqrt{s}$ and $\mttbar$ has been 
estimated in detail in~\cite{Beccaria:2004sx}, and should anyhow be scarcely relevant if a 
complete integration (i.e. from threshold to the "end" point $\mttbar\simeq 1.2$ TeV 
is performed).

Eq.~(\ref{eq:number}) defines the number of top-antitop pairs with a certain given 
helicity that
are produced in the energy interval $(\sqrt{s_1}, \sqrt{s_2})$, ignoring the precise details 
of the invariant mass distributions. From these expressions we can now express 
the differences that would correspond to the original Eqs.~(\ref{eq:diff1}-\ref{eq:diff2}), i.e.:

\ba
\label{eq:nll-nrr}
\lefteqn{N_{LL}(s_1, s_2)-N_{RR}(s_1, s_2) =} && \\
&&  {\cal L}_{int}\int_{s_1}^{s_2} ds \ {1\over S}~\int^{\cos\vartheta_{max}}_{\cos\vartheta_{min}}
d\cos\vartheta\ L\left(\frac{s}{S}, \cos\vartheta\right)\ 
\frac{d\sigma^{\rm Born}_{LL=RR}}{d\cos\vartheta}\ \alpha\ (F_{LL}^{\rm EW} - F_{RR}^{\rm EW})
\nonumber
\ea

\ba
\label{eq:nlr-nrl}
\lefteqn{N_{LR}(s_1, s_2)-N_{RL}(s_1, s_2) =} && \\
&&  {\cal L}_{int} \int_{s_1}^{s_2} ds \ {1\over S}~\int^{\cos\vartheta_{max}}_{\cos\vartheta_{min}}
d\cos\vartheta\ L\left(\frac{s}{S}, \cos\vartheta\right)\ 
\frac{d\sigma^{\rm Born}_{LR=RL}}{d\cos\vartheta}\ \alpha\ (F_{LR}^{\rm EW} - F_{RL}^{\rm EW})
\nonumber
\ea

One sees that now these (more realistic) sum rules are not completely free of 
QCD effects, since the latter are implicitly affecting the used value of the 
various parton distribution functions. This seems to be a price to pay to the 
purpose of a meaningful experimental verification of the sum rules. However, 
one may also assume the consistency philosophy of the paper, i.e. that the 
relevant parton distribution functions (in this case, the gluon-gluon ones)
should be estimated at Born level, or alternatively that they will be independently 
derived by 
other measurements. In this sense, the QCD independence of Eqs.~(\ref{eq:nll-nrr},\ref{eq:nlr-nrl}) 
would  be valid not only for the virtual effects of the final states 
diagrams, but for the overall process.

The question that should now be asked is that of whether an experimental 
analysis of the realistic experimental errors on the differences that appear 
in Eqs.~(\ref{eq:nll-nrr},\ref{eq:nlr-nrl}) does or will exist. 
At the moment, we are not aware of 
such an investigation, that appears to us reasonably well motivated and 
auspicable. The only experimental measurement that appears to have been 
considered with sufficient details is that of a different asymmetry, 
originally called $C$ in~\cite{Stelzer:1995gc} and very recently reconsidered in some papers~\cite{C}.
In our notation, we would define it as :

\be
\label{eq:Cdef}
C(s_1, s_2) = \frac{(N_{LL}+N_{RR})-(N_{LR}+N_{RL})}{(N_{LL}+N_{RR})+(N_{LR}+N_{RL})}, 
\qquad N_{ab}\equiv N_{ab}(s_1, s_2)
\ee

For the latter quantity, a recent estimate~\cite{Sonnenschein}
proposes the value (assuming 30 fb$^{-1}$ statistics and integrating over the full energy range) :
\be
\label{eq:Cval}
C = 0.311^{+0.034}_{-0.035}\ ({\rm stat})\pm 0.028\ ({\rm syst})
\ee

A priori, the result (\ref{eq:Cval}) cannot be directly related to our Eqs.~(\ref{eq:nll-nrr},\ref{eq:nlr-nrl}). 
One sees that the proposed value for $C$ implies a determination 
of the  two "blocks " of opposite helicity (LL+RR) and equal helicity (LR + 
RL) pairs, without selecting within the blocks the fraction of left-handed (or 
right- handed) top components, and analogously for the equal helicity case. 
Waiting for a dedicated experimental analysis of our suggestion, we have tried 
to produce some qualitative numbers to be compared e.g. with that of Eq.~(\ref{eq:Cval}). 
This analysis, that we consider a purely indicative one, is 
illustrated in the final forthcoming Section~(\ref{sec:numerics}).

\section{Tentative Numerical Analysis}
\label{sec:numerics}

To produce some numerical results, we do not have many reliable possibilities 
at this stage of our theoretical investigations. In practice, the only 
possible example that we can propose is that of an estimate of the first of 
the two candidate asymmetries, i.e. Eq.~(\ref{eq:nll-nrr}). 
Assuming the MSSM theoretical 
description, we have reasons to believe, from the investigation performed in~\cite{Beccaria:2004sx}, 
that in the so called "reasonably light SUSY scenario", where all SUSY masses lie below ~400 GeV, 
a logarithmic expansion of Sudakov kind, computed at NLO, should provide a valid description of the 
electroweak one-loop effects of the model, in the $\sqrt{s}$ region from $\sim~0.7$ to 
$\sim~~1.3$ TeV (on the contrary, we do not have at disposal a similar simple 
expansion for the remaining LR, RL amplitudes, that would require a complete 
calculation, not yet available). In this spirit, we have thus estimated the 
two numbers $N_{LL}$ and $N_{RR}$ entering the difference Eq.~(\ref{eq:nll-nrr}) with 
\be
\sqrt{s_1} = M_{t\overline t, \rm min}, \qquad \sqrt{s_2} = 1.2\ \mbox{TeV},
\ee
and ${\cal L}_{int} = 30 \mbox{fb}^{-1}$. The upper limit has been kept fixed. The lower limit
$M_{t\overline t, \rm min}$ has been varied in the analysis.
For the parton distribution functions 
we have used the lastest MRST set~\cite{wlumi}, although we insist that in 
this way a part of the ${\cal O}(\alpha\alpha_s)$ corrections has been inserted, that 
could be considered as a future experimental input. 

\begin{figure}
\centering
\epsfig{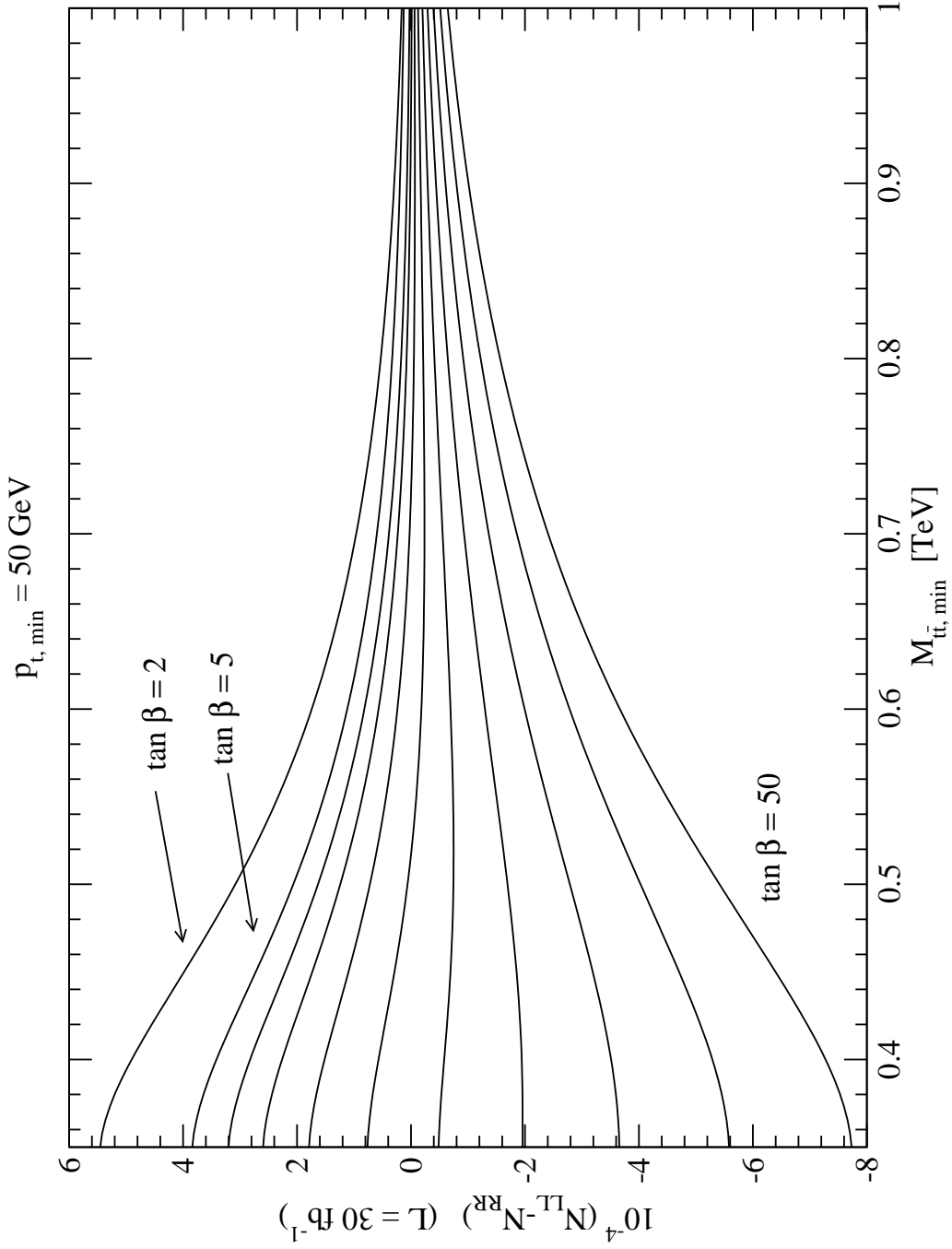}
\vspace{1.5cm}
\caption{
Difference $N_{LL}-N_{RR}$ computed with ${\cal L}_{int} = 30\mbox{fb}^{-1}$ and plot as
a function of the lower limit $M_{t\overline t, \rm min}$ in the energy integration. The various
curves are associated to different $\tan\beta$ (2, 5, 10 , 15, \dots from top to bottom).
}
\label{fig:LL-RR}
\end{figure}

\begin{figure}
\centering
\epsfig{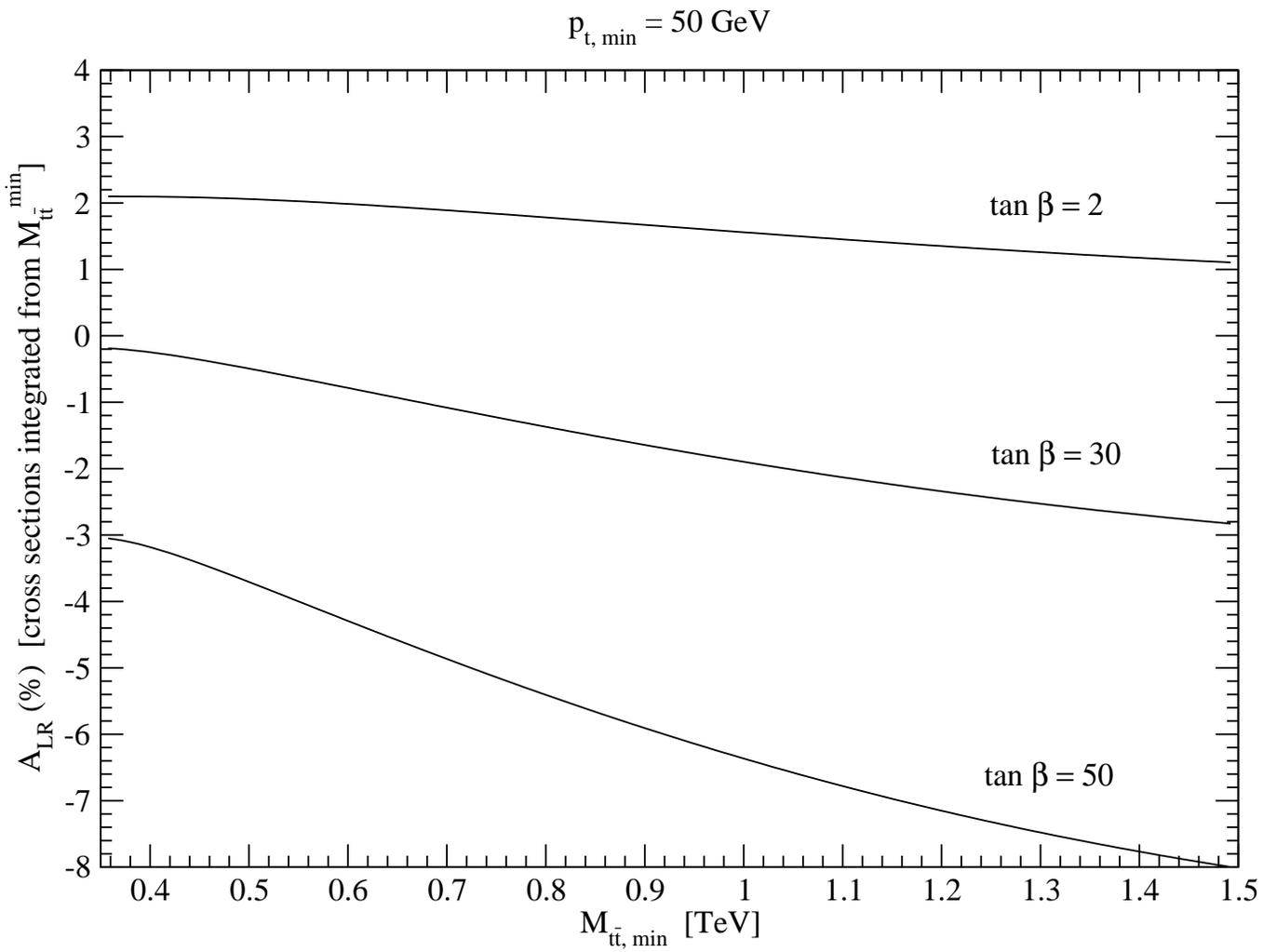}
\vspace{1.5cm}
\caption{
Asymmetry $(N_{LL}-N_{RR})/(N_{LL}+N_{RR})$ with variable lower limit $M_{t\overline t, \rm min}$ in the energy integration. 
}
\label{fig:ALR}
\end{figure}

The result of our calculation are shown in Figs.~(\ref{fig:LL-RR}) and (\ref{fig:ALR}).
As one sees, the lower limit of integration has been 
allowed, for purely illustrative purposes, to reach the physical threshold of 
the process, where our assumed asymptotic expansion is very unlikely to be 
valid. For a reasonably meaningful interpretation of our analysis, one should
consider a lower limit of $\mttbar$ of at least, say, $700-800$~GeV. In this 
range, our results are plotted as a function of $\tan\beta$, that is the only free 
parameter in the assumed MSSM scenario entering the NLO logarithmic expansion 
(the light SUSY scale would enter at next-to next leading order). 
Keeping in mind the limitations of our analysis, we mention some features that 
seem to us to be, least to say, encouraging. First of all, when $\tan\beta$ varies 
in its considered range (2-50), Fig.~(\ref{fig:LL-RR}) shows that the difference of the number 
of pairs changes sign, moving from a positive value of $\sim+1.5\cdot 10^4$ to a 
negative one $\sim-3.5\cdot 10^4$, at the assumed luminosity. This fact can be 
expressed in terms of the values of the integrated asymmetry, as shown in 
Fig.~(\ref{fig:ALR}).
Here one sees a variation of this quantity from $\sim +2\%$ percent to $\sim-5\%$ in 
the $\tan\beta$ range. Perhaps more relevant is the fact that, varying $\tan\beta$ in 
the large values range, from  30 to 50, the asymmetry varies from $\sim-1\%$
to $\sim-5\%$. This implies that a measurement of the integrated asymmetry 
performed with a {\em hundred percent precision}, leading to a central value 
within this range, could lead to a valuable discrimination of candidate 
$\tan\beta$ values, to be combined with other existing $\tan\beta$ determination 
proposals~\cite{TanBeta}. Clearly, a more rigorous calculation, valid 
in a less special scenario and extended to the remaining sum rule, would be 
requested. This is in fact being performed at the moment~\cite{Moretti}.

To conclude this paper, we add another short proposal. We have derived until 
now the two sum rules Eqs.~(\ref{eq:nll-nrr}-\ref{eq:nlr-nrl}), and one major difficulty has been 
represented by the lack of a dedicated experimental analysis on the proposed 
quantities. Since, on the contrary, an experimental estimate exists for the 
quantity C defined in Eq.~(\ref{eq:Cdef}), one might wonder whether some possible information 
could be obtained combining the latter Eq.~(\ref{eq:Cval}) with our proposed sum rules 
Eqs.~(\ref{eq:nll-nrr}-\ref{eq:nlr-nrl}). We want to show that, working consistently in the one-loop 
approximation, this is actually the case. With this purpose, we shall make the 
extra assumption (that appears rather natural to us) that a precise 
experimental information exists also on the total number of top-antitop pairs, 
defined in our notations as :

\be
N_T = N_{LL}+N_{RR}+N_{LR}+N_{RL},
\ee
and from now on we do not include in the notations the limits $(s_1, s_2)$ which 
are the same as in our previous discussions, {\em i.e.} $N_{ab}\equiv N_{ab}(s_1, s_2)$.
In this spirit, we write now the general expressions for the separate $N_{ab}$
\ba
\label{eq:general}
\lefteqn{N_{ab} = {\cal L}_{int}\int_{s_1}^{s_2} ds \ {1\over S}~\int^{\cos\vartheta_{max}}_{\cos\vartheta_{min}} 
d\cos\vartheta\ L\left(\frac{s}{S}, \cos\vartheta\right)\ \times} && \\
&& \qquad \qquad \times \qquad \frac{d\sigma^{\rm Born}_{ab}}{d\cos\vartheta}\ (1+\alpha_s\ F_{ab}^{\rm QCD} + \alpha\ F_{ab}^{\rm EW}). \nonumber
\ea
In the four cases $N_{ab}$, $ab = LL, LR, RL, RR$, there are actually only 2 independent QCD correction factors.
Note that, in fact, this is the point of our analysis, since we have shown that on general grounds 
the QCD correction is the same for the pair (LL, RR) or for the pair (LR, RL).
Therefore, we can write Eq.~(\ref{eq:general}) as 
\ba
N_{LL} &=& N\ +\alpha_s\ \delta^{\rm QCD}_1+\alpha\ \delta^{\rm EW}_{LL},  \\
N_{RR} &=& N\ +\alpha_s\ \delta^{\rm QCD}_1+\alpha\ \delta^{\rm EW}_{RR},  \\
N_{LR} &=& N'\ +\alpha_s\ \delta^{\rm QCD}_2+\alpha\  \delta^{\rm EW}_{LR}, \\
N_{RL} &=& N'\ +\alpha_s\ \delta^{\rm QCD}_2+\alpha\  \delta^{\rm EW}_{RL},
\ea
where $N$ and $N'$ are the Born values $N_{LL}^{Born}=N_{RR}^{Born}$ and $N_{LR}^{Born}=N_{RL}^{Born}$, respectively,
and $\delta^{\rm QCD}_{1,2}$, $\delta^{\rm EW}_{ab}$ can be derived from Eq.~(\ref{eq:general}).
Since only two QCD theoretical quantities appear, we can at this point bargain 
them in terms of the measured quantities $C$, $N_T$.
This leads us to the following 4 equations:
\ba
\label{eq:final:begin}
N_{LL} &=& \frac{1}{4}N_T(1+C)+\frac{1}{2}\ \alpha\ (\delta_{LL}^{\rm EW}-\delta_{RR}^{\rm EW}), \\
N_{RR} &=& \frac{1}{4}N_T(1+C)-\frac{1}{2}\ \alpha\ (\delta_{LL}^{\rm EW}-\delta_{RR}^{\rm EW}), \\
N_{LR} &=& \frac{1}{4}N_T(1-C)+\frac{1}{2}\ \alpha\ (\delta_{LR}^{\rm EW}-\delta_{RL}^{\rm EW}), \\
\label{eq:final:end}
N_{RL} &=& \frac{1}{4}N_T(1-C)-\frac{1}{2}\ \alpha\ (\delta_{LR}^{\rm EW}-\delta_{RL}^{\rm EW}), 
\ea
One sees that the previous equations allow to express each separate possible 
helicity pair production in terms of $N_T$, $C$ and of 4 purely weak 
one-loop corrections, to be estimated theoretically once the candidate model 
is fixed. 

Eqs.~(\ref{eq:final:begin}-\ref{eq:final:end})
are the most general expressions of the considerations of our paper. 
They relate observable quantities of the process, with all possible polarization properties, 
to purely weak effects, and would therefore provide a clean test of the genuinely weak sector of 
candidate theoretical models. In our opinion, they would deserve a dedicated experimental investigation. 
This was in fact the main goal of our paper, and we hope that it will inspire fruitful 
practical consequences.

\newpage

\newpage

\newpage

\end{document}